\documentstyle[preprint,aps]{revtex}
\tighten
\draft
\begin{document}

\title{\bf Collective excitations in the inner crust of neutron stars :
supergiant resonances}

\author{\rm E. Khan$^{a}$, N. Sandulescu$^{a,b,c}$ and Nguyen Van Giai$^{a}$}

\address{a) Institut de Physique Nucl\'eaire, Universit\'e Paris-Sud, IN$_{2}$P$_{3}$-CNRS,
91406 Orsay Cedex, France\\
b) Institute of Physics and Nuclear Engineering, 76900 Bucharest, Romania \\
c) Department of Physics, University of Washington, Seattle, WA 98195, USA}

\maketitle

\begin{abstract}
We investigate the nuclear collective excitations of Wigner-Seitz cells
containing nuclear clusters immersed in a gas of neutrons. This baryonic
non-uniform system is specific to the structure of inner crust matter of
neutron stars. The collective excitations are studied in the framework of a
spherical Hartree-Fock-Bogoliubov + Quasiparticle Random Phase
Approximation, formulated in coordinate representation. The calculations are
done for two representative Wigner-Seitz cells with baryonic density equal
to 0.02 fm$^{-3}$ and 0.08 fm$^{-3}$. It is shown that the excitations with
low multipolarities are concentrated almost entirely in one strongly
collective mode which exhausts a very large fraction of the energy-weighted
sum rule. Since these collective modes are located at very low energies
compared to the giant resonances in standard nuclei, they may affect
significantly the specific heat of baryonic inner crust matter of neutron
stars.
\end{abstract}

\vskip 0.5cm
{\it PACS numbers}: 21.60Jz, 26.60+c, 24.30Cz, 21.10Re


\newpage

The cooling of low-mass neutron stars is influenced strongly by the
 superfluid properties of inner crust matter \cite{ya04}. The superfluid
 properties and their influence on the specific heat were analyzed rather
 intensively in the past years by using various frameworks, e.g. semiclassical
 pairing models \cite{br94,el96}, Bogoliubov-type calculations based on a
 Woods-Saxon mean field \cite{ba98,pi02}, and the self-consistent
 Hartree-Fock-Bogoliubov (HFB) approach \cite{sa04,sa04b}. These
 calculations showed that the specific heat of baryonic inner crust matter
 can be reduced by orders of magnitudes due to pairing correlations.

  In all the calculations mentioned above the specific heat was evaluated by
  considering only non-interacting quasiparticles states. However, the
  specific heat can be also strongly affected by the collective modes
  created by the residual interaction between the quasiparticles, especially
  if these modes appear at low-excitation energies. To our knowledge, at
  present there is not any microscopic study of such nuclear collective modes
  induced in the inner crust matter of neutron stars. The scope of this
  Letter is to present such an investigation. The specific heat of the inner
  crust is also largely determined by the electrons motion and, to a lesser
  extent, by the lattice vibrations \cite{ya04,pi02,pe95,ma04}. These degrees
  of freedom of the inner crust matter are not discussed here.

 In microscopic calculations the inner crust matter is usually treated in
the Wigner-Seitz (WS) approximation \cite{pe95,ne73}. Accordingly the inner
crust is modelised by non-interacting cells containing a neutron-rich
nucleus immersed in a dilute gas of neutrons and relativistic electrons. For
baryonic densities ranging from 1.4 10$^{-3}$$\rho_0$ to about 0.5$\rho_0$,
where $\rho_0$=0.16 fm$^{-3}$ is the nuclear matter saturation density, the
nuclear clusters are considered spherical \cite{ne73,do00}. At higher
densities, the inner crust matter can develop various non-spherical phases
(e.g.,rods, slabs, tubes, bubbles) \cite{pe95}. In this Letter we will
focus on the nuclear collective modes developed in WS cells containing
spherical nuclear clusters. For illustration we will consider two
representative WS cells chosen from Ref.\cite{ne73}, i.e., one formed by 50
protons and 1750 neutrons and an other formed by 32 protons and 950
neutrons. These two cells will be quoted below as $^{1800}$Sn and
$^{982}$Ge. The density of the neutron gas far from the nuclear cluster is
of about 0.018 fm$^{-3}$ in the first cell and about 0.074 fm$^{-3}$ in the
second. The radius of the first (second) cell is 27.6 (14.4) fm.

 The collective response is calculated in the HFB+Quasiparticle Random Phase
 Approximation (QRPA) approach formulated in the coordinate representation
 \cite{kh02}. In the first step of the calculation we solve the HFB
 equations for the ground state of the given Wigner-Seitz cell, considered
 as an isolated system. The HFB calculations are performed in coordinate
 representation and imposing Dirichlet-Neuman boundary conditions at the
 edge of the cell \cite{ne73}. The effective forces used in the HFB
 calculations are the same as in Ref.\cite{sa04b}, i.e., the Skyrme
 interaction SLy4 \cite{ch98} in the particle-hole channel and a
 density-dependent contact force for the pairing interaction. The parameters
 of the pairing interaction are chosen so as to obtain in neutron matter
 approximatively the same pairing gap as given by the Gogny force
 \cite{sa04,be91,ga99}. The HFB results for the pairing fields and particle
 densities of the two WS cells mentioned above are shown in Figures 1 and 2.

The solutions of the HFB equations are used in the next step to construct
the QRPA Green function. In the Green function are introduced all the
quasiparticle states with energies up to 60 MeV, which is the cut-off energy
employed in the HFB calculations. Both HFB and QRPA equations are solved on
a radial mesh with a 0.2 fm step and the Green function is calculated by
using an energy smoothing factor equal to 150 keV. The residual interaction
between the quasiparticles is calculated from the second derivative of the HFB
energy functional with respect to the particle and pairing densities. For
calculating the residual interaction in the particle-hole channel we apply
the Landau-Migdal approximation \cite{ba75}, which simplifies the numerical
treatment of the velocity-dependent terms of the Skyrme interaction. Due to
this approximation the full self-consistency of HFB+QRPA calculations is
slightly broken. As a result, the spurious mode associated to the
translational invariance is not located exactly at zero energy. In order to
bring it to zero we renormalise the particle-hole part of the residual
interaction by a factor of 0.93. The Coulomb and the spin-orbit residual
interactions between quasiparticles are dropped since they play a minor role
compared to the other interactions. However, both the Coulomb and spin-orbit
interactions are included in the HFB mean-field calculations. Further details
on how the collective response is calculated in the HFB+QRPA approach can
be found in Refs.\cite{kh02,kh04}.

In order to analyse the microscopic content of the response function we need
to express the forward (X) and backward (Y) amplitudes of the QRPA operators
\cite{ri80} in coordinate representation. Thus it can be shown that the
forward amplitudes corresponding to an exciting state $\nu$ have the
folowing form:

\begin{equation}\label{eq:x}
 	\begin{array}{c}
 X^{\nu}_{ij}=\frac{1}{2}\int d{\bf r}\sum_{\sigma}\rho^{\nu}({\bf r})
	\left(V^{*}_{j}\left({\bf r}\sigma\right)
	      U^{*}_{i}\left({\bf r}\sigma\right) -
	      V^{*}_{i}\left({\bf r}\sigma\right)
	      U^{*}_{j}\left({\bf r}\sigma\right)\right) +
\\
	\kappa^{\nu}({\bf r})
	     U^{*}_{j}\left({\bf r}\bar{\sigma}\right)
	     U^{*}_{i}\left({\bf r}\sigma\right)    +
	\bar{\kappa}^{\nu}({\bf r})
	     V^{*}_{j}\left({\bf r}\sigma\right)
	     V^{*}_{i}\left({\bf r}\bar{\sigma}\right)
\\
\end{array}
\end{equation}
where $\rho^{\nu}$ is the particle-hole transition density for the state
$\nu$ while $\kappa^{\nu}$ and ${\bar \kappa}^{\nu}$ are the pair addition
and pair removal transition densities \cite{kh04}. The quantities $U_i(r\sigma)$
and $V_i(r\sigma)$ are the two components of the HFB wave functions and
$\sigma$ is the spin index. The time reversed wave functions are denoted
by the spin index $\bar{\sigma}$.

Within the HFB+QRPA formalism presented above we first calculate the
quadrupole neutron response in the cell $^{1800}$Sn. The corresponding
transition operator is r$^2$Y$_{20}$. The unperturbed HFB response, built by
non-interacting quasiparticle states, and the QRPA response are shown in
Figure 3. As can be clearly seen, when the residual interaction is
introduced among the quasiparticles the unperturbed spectrum, distributed
over a large energy region, is gathered almost entirely in a the peak
located at about 3 MeV. This peak collects more than 99$\%$ of the total
quadrupole strength and it exhausts about 70$\%$ of the energy-weighted sum
rule. This mode is extremely collective. Thus, there are more than one
hundred two-quasiparticle configurations which contribute to this mode, with
the two main configurations contributing no more than 5\%. Another
indication of the extreme collectivity of this low-energy mode can be seen
from its reduced transition probability, B(E2), which is equal to 25.10$^3$
Weisskopf units. This value of B(E2) is two orders of magnitude higher than
in standard nuclei. We notice also that an extrapolation of the energy
position based on the GDR systematics in finite nuclei, i.e., 65.A$^{-1/3}$
MeV \cite{wo87}, would predict the low-energy peak at about 5 MeV. This
underlines the fact that this WS cell cannot be simply considered as a giant
nucleus. As discussed below, the reason is that in this cell the collective
dynamics of the neutron gas dominates over the cluster contribution. Apart
from the quadrupole mode discussed above, we have also investigated the
response of the WS cell to the dipole and monopole excitations: they
display similar features, leading to the same qualitative conclusions.

  It is interesting to examine what would be the energy of the
  Bogoliubov-Anderson (BA) excitation mode \cite{bo59,an58} which the
  neutron superfluid gas would develop in the WS cell. For a low-momentum
  $p$ the energy of the BA mode is given approximatively by $\omega= v_F
  p/\sqrt3$, where $v_F$ is the Fermi velocity. To estimate the BA mode in a
  WS cell one can use for $p$ the values provided by the condition
  $j_\lambda (kR)$=0, where $j_\lambda$ is the spherical Bessel function,
  $\lambda$ is the multipolarity of the excitation, R is the radius of the
  cell, and $k=p/\hbar$. The value of $v_F$ is obtained from the density of
  the neutron gas, taking the mass of neutrons equal to the bare mass. Using
  these approximations one gets for the lowest quadrupole mode an energy
  equal to 3.1 MeV, which is close to the energy extracted from the HFB+QRPA
  calculations.  Thus the low-lying quadrupole excitations in the cell
  $^{1800}$Sn can be viewed as hydrodynamic sound-type modes produced by the
  neutron superfluid. The hydrodynamic picture is supported by the small
  coherence length of neutron gas superfluid, of about 3 fm, compared to the
  dimension of the WS cell. The value of the coherence length quoted above
  corresponds to the approximation $\xi=\hbar v_F/(\pi \Delta)$ \cite{ge89},
  where $\Delta$ is the averaged pairing gap in the neutron gas region.

 In order to study how the collective excitations would behave at higher
 baryonic densities of inner crust matter, we have also calculated the
 response for the cell $^{982}$Ge. As seen in Figs.1-2, in this cell the
 cluster and the neutron gas are less distinct than in the cell $^{1800}$Sn.
 Therefore one expects that the collective modes developed in the cell
 $^{982}$Ge are rather different from the hydrodynamic modes of the neutron
 gas superfluid.

 The neutron quadrupole response of $^{982}$Ge provided by the HFB+QRPA
 calculations is shown in Figure 4. One of most striking feature seen in
 Figure 4 is the huge difference between the unperturbed and QRPA response.
 Thus one can see that by introducing the residual interaction between
 quasiparticle states, the unperturbed strength, dominated by the peak at 11
 MeV, is shifted down and distributed through an energy region of about 9 MeV.
 The lowest quadrupole mode appears at 1.8 MeV. This mode is built in
 proportion of $23\%$ by the two-quasiparticle configurations (15$_{31/2}$
 17$_{35/2}$) and (15$_{29/2}$ 17$_{33/2}$). In these configurations, which
 form also the main part of the unperturbed spectrum, the single-particle
 states corresponding to the first (second) quasiparticle states are almost
 entirely occupied (empty). Due to the large degeneracy of these states, the
 residual interaction forms many particle-hole configurations which add
 coherently to the low-energy quadrupole mode. As expected, due to the
 dominance of these particle-hole configurations, the main features of the
 low-lying peak will be present also in a HF+RPA calculation. This is indeed
 the case, as seen in the upper part of Figure 4. It should be however noted
 that the pairing residual interaction significantly spreads the strength of
 the RPA peak located above 8 MeV.

In conclusion, the collective response of two representative WS cells has
been calculated fully microscopically, using the HFB + QRPA approach. We
found that in both Wigner-Seitz cells the residual interaction between the
quasiparticle states generates a sort of supergiant resonances located at
low energies. These energies have the same order of magnitude than the average
pairing gap of neutron superfluid. Consequently, these collective modes can
affect significantly the entropy and the specific heat of baryonic inner
crust matter. A quantitative estimation of these effects would require
finite-temperature HFB+QRPA calculations
\cite{kh04b}. These calculations will be the subject of a future study.

\vskip 0.1cm
 One of us (NS) acknowledges valuable discussions with A.Bulgac.

\newpage\eject
\noindent
{\Large \bf Figure captions}
\vskip 48pt
\par\noindent
{\bf Figure 1.} Particle densities calculated with HFB for the WS cells
$^{1800}$Sn (solid line) and $^{982}$Ge (dashed line). \\ \\
{\bf Figure 2.} Pairing field calculated with HFB for the WS cells
$^{1800}$Sn (solid line) and $^{982}$Ge (dashed line). \\\\
{\bf Figure 3.} Quadrupole strength distribution of neutrons for the cell $^{1800}$Sn.
The full curve represents the QRPA strength and the dashed line is the
HFB unperturbed strength. \\\\
{\bf Figure 4.} The same as in Figure 3 but for the cell $^{982}$Ge. The inset
shows the response calculated in the HF+RPA approach. \\\\

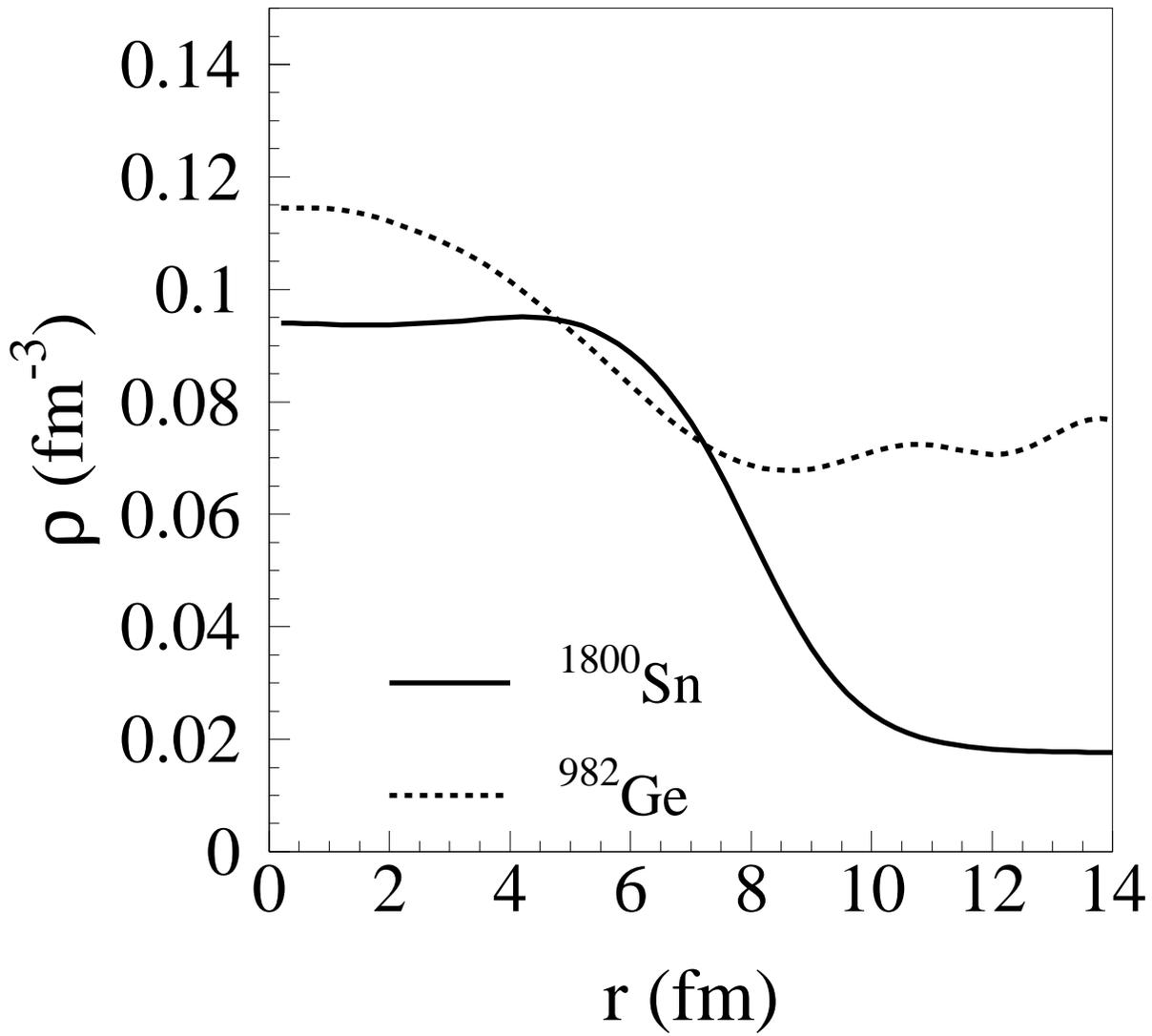
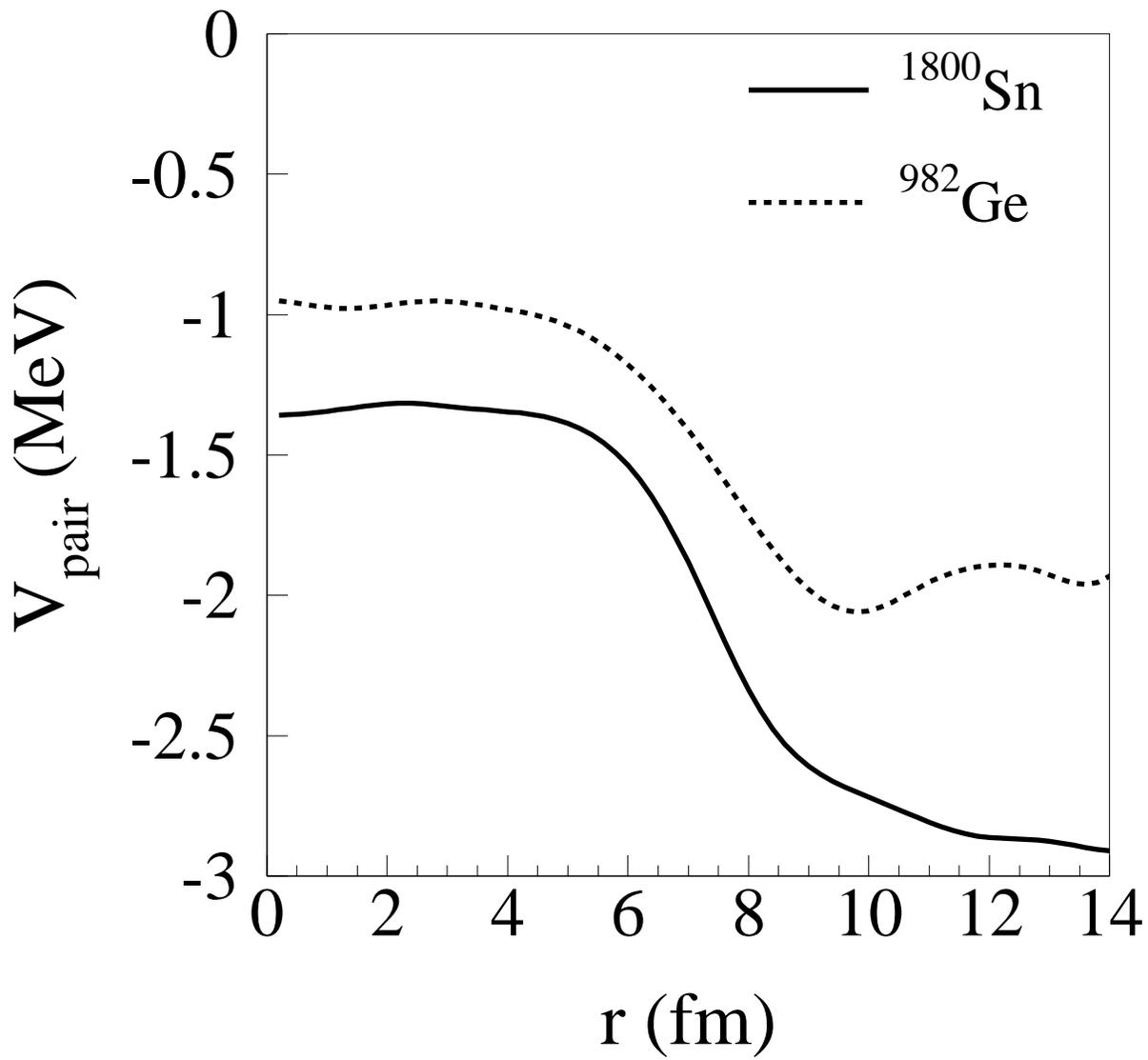
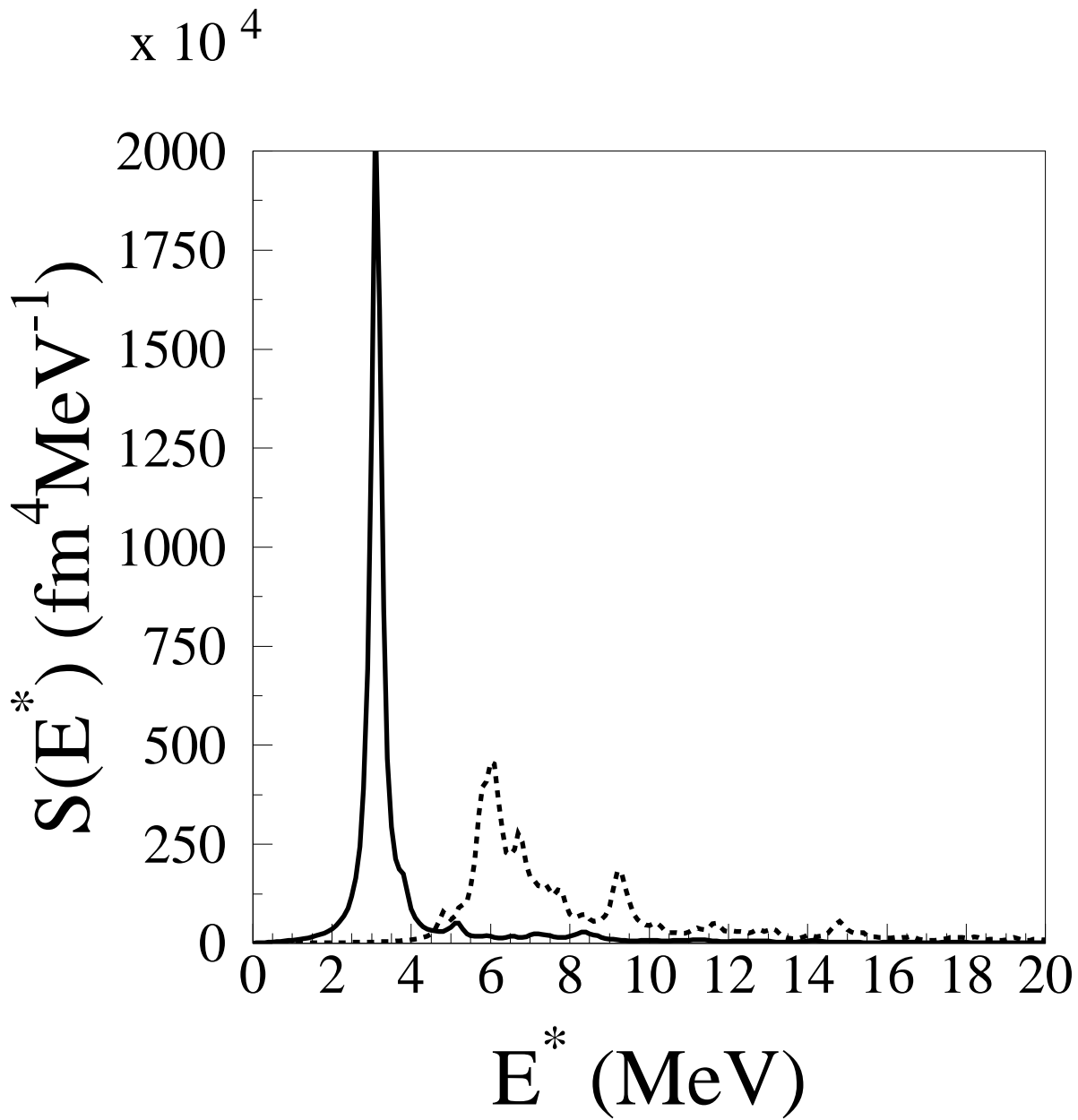
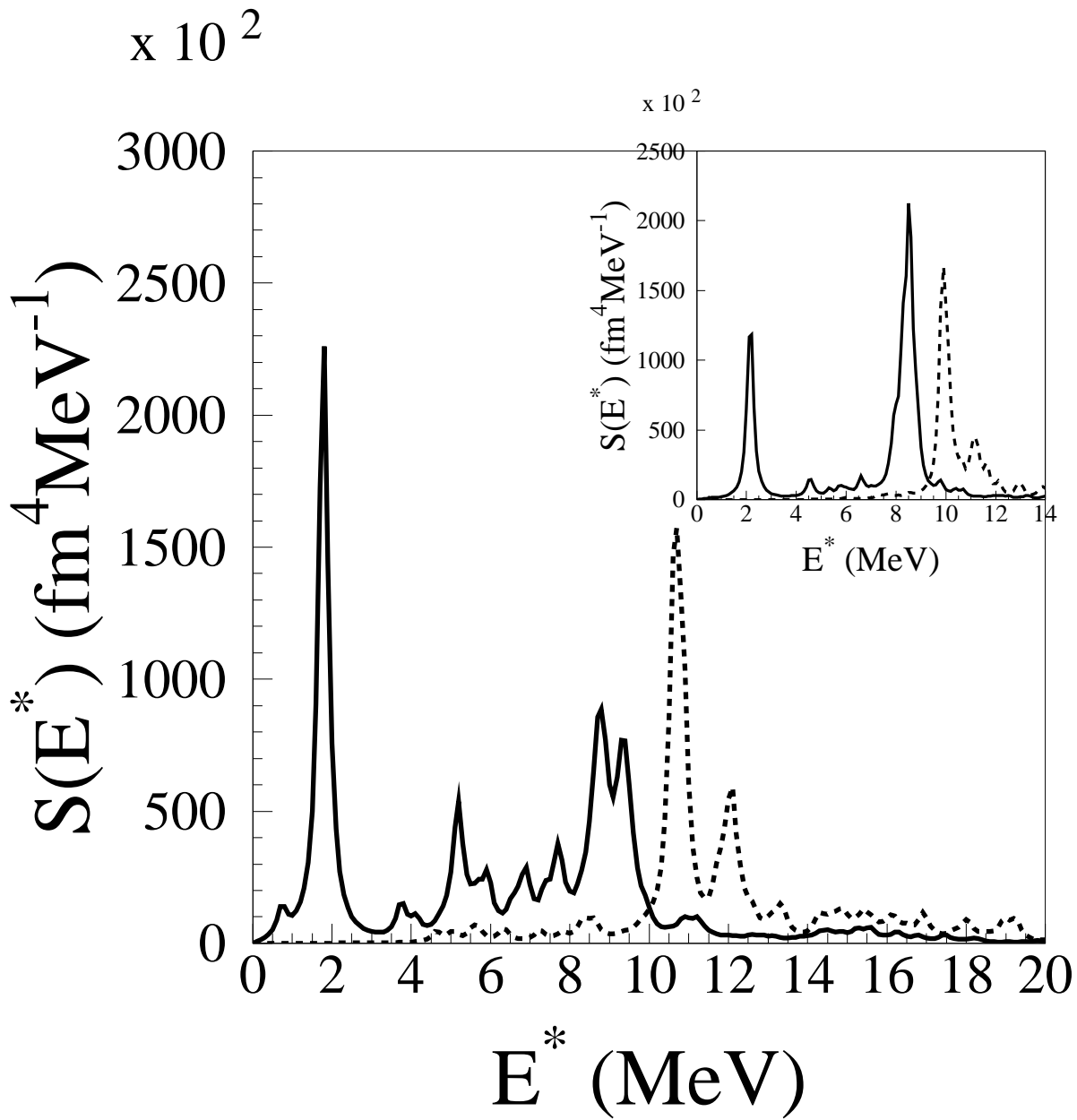
\end{document}